\documentclass[conference]{IEEEtran}
\IEEEoverridecommandlockouts
\usepackage{cite}
\usepackage{amsmath,amssymb,amsfonts}
\usepackage{algorithmic}
\usepackage{graphicx}
\usepackage{textcomp}
\usepackage{xcolor}
\usepackage{multirow}
\usepackage{booktabs}
\def\BibTeX{{\rm B\kern-.05em{\sc i\kern-.025em b}\kern-.08em
    T\kern-.1667em\lower.7ex\hbox{E}\kern-.125emX}}
\begin{document}

\title{A Study on Zero-shot Non-intrusive Speech Assessment using Large Language Models}

\author{\IEEEauthorblockN{Ryandhimas E. Zezario}
\IEEEauthorblockA{
\textit{Academia Sinica}\\
Taipei, Taiwan \\
ryandhimas@citi.sinica.edu.tw}
\and
\IEEEauthorblockN{Sabato M. Siniscalchi}
\IEEEauthorblockA{
\textit{University of Palermo}\\
Palermo, Italy \\
sabatomarco.siniscalchi@unipa.it}
\and
\IEEEauthorblockN{Hsin-Min Wang}
\IEEEauthorblockA{
\textit{Academia Sinica}\\
Taipei, Taiwan \\
whm@iis.sinica.edu.tw}
\and
\IEEEauthorblockN{Yu Tsao}
\IEEEauthorblockA{
\textit{Academia Sinica}\\
Taipei, Taiwan \\
yu.tsao@citi.sinica.edu.tw}

}

\maketitle

\begin{abstract}
This work investigates two strategies for zero-shot non-intrusive speech assessment leveraging large language models. First, we explore the audio analysis capabilities of GPT-4o. Second, we propose GPT-Whisper, which uses Whisper as an audio-to-text module and evaluates the text’s naturalness via targeted prompt engineering. We evaluate the assessment metrics predicted by GPT-4o and GPT-Whisper, examining their correlation with human-based quality and intelligibility assessments and the character error rate (CER) of automatic speech recognition. Experimental results show that GPT-4o alone is less effective for audio analysis, while GPT-Whisper achieves higher prediction accuracy, has moderate correlation with speech quality and intelligibility, and has higher correlation with CER. Compared to SpeechLMScore and DNSMOS, GPT-Whisper excels in intelligibility metrics, but performs slightly worse than SpeechLMScore in quality estimation. Furthermore, GPT-Whisper outperforms supervised non-intrusive models MOS-SSL and MTI-Net in Spearman’s rank correlation for Whisper's CER. These findings validate GPT-Whisper's potential for zero-shot speech assessment without requiring additional training data.
\end{abstract}

\begin{IEEEkeywords}
speech assessment, zero-shot, non-intrusive, whisper, ChatGPT, large language model
\end{IEEEkeywords}

\section{Introduction}
Speech assessment metrics play a critical role in evaluating a variety of speech-related applications, including speech enhancement \cite{loizou2007speech, ref_19}, hearing aid (HA) devices \cite{katehaspi, kates2014hearingb, chiang2023multiobjective}, and telecommunications \cite{polqa_2013}. With the advancement of deep learning and the need for accurate non-intrusive speech assessment metrics, researchers have increasingly adopted deep learning models for speech assessment \cite{yang22o_interspeech, 10447907, mogridge2024nonintrusive, candy, zezario2024studyincorporatingwhisperrobust, manocha,wang2024enablingauditorylargelanguage,cooper2024review, fu2024selfsupervised}. To achieve accurate automatic assessments, various strategies have been explored, such as reducing listener bias \cite{ldnet2022}, integrating large pre-trained models (e.g., self-supervised learning (SSL) models \cite{ssl-mos, mosa-net}, Whisper \cite{zezario2024studyincorporatingwhisperrobust}, and speech language models\cite{speechlm}), utilizing ensemble learning \cite{yang22o_interspeech, 10447907, mogridge2024nonintrusive}, and incorporating pseudo labels \cite{zezario2023multitask}. Despite significant performance improvements, achieving satisfactory generalization with limited training samples remains challenging. 

Recently, with the development of large-scale conversational agents like ChatGPT, there has been an increasing interest in evaluating their ability to perform more extensive reasoning and understanding \cite{wei2024chatiezeroshotinformationextraction}. The capabilities of GPT-4, especially its latest extension GPT-4o \footnote{https://platform.openai.com/docs/models/gpt-4o}, have been significantly expanded to enable not only advanced text understanding but also multimodal integration, such as merging visual and audio understanding. In the case of image understanding, ChatGPT has been successfully integrated with visual language models to perform tasks such as deep fake detection \cite{Jia_2024_CVPR}. For audio understanding, a noteworthy integration is AudioGPT \cite{Huang_Li_Yang_Shi_Chang_Ye_Wu_Hong_Huang_Liu_Ren_Zou_Zhao_Watanabe_2024}. This approach integrates ChatGPT with multiple pre-trained audio models. Based on specific prompt input, the most appropriate audio model is selected to generate the response. While AudioGPT excels at integrating audio understanding with prompt engineering, it does not cover speech assessment. Given the interest in reliable non-intrusive speech assessment with minimal training samples, we intend to investigate whether ChatGPT can effectively perform speech assessment in a zero-shot setting and identify optimal strategies to ensure accurate and unbiased results.

In this paper, we explore two strategies for zero-shot non-intrusive speech assessment by leveraging large language models (LLM). First, we directly leverage the audio analysis capabilities of GPT-4o for speech assessment. Second, we propose a more advanced approach, namely GPT-Whisper. Specifically, we use Whisper \cite{Whisper} as an audio-to-text module to create text representations, which are then assessed using targeted prompt engineering focused on evaluating the naturalness of the predicted text. To gain a deeper insight into our new assessment metric computed by GPT-Whisper, we also evaluate its correlation with several metrics, including human-based quality assessment, human-based intelligibility assessment, and the character error rate (CER) of automatic speech recognition (ASR) models. We use two open-source ASR models: Whisper \cite{Whisper} and Google ASR \cite{ASR} to generate CER scores. To the best of our knowledge, this is the first attempt to leverage ChatGPT to simultaneously estimate speech quality, intelligibility, and CER. Finally, we compare our approach with DNSMOS \cite{dnsmos}, one unsupervised neural speech assessment model—SpeechLMscore \cite{speechlm}—and two supervised neural speech assessment models—MOS-SSL \cite{ssl-mos} and MTI-Net \cite{zezario2022mti}.

Our experimental results first confirm that using ChatGPT alone for audio analysis is less optimal because it mainly relies on factors such as amplitude range, standard deviation, and signal-to-noise ratio (SNR) to assess the quality or intelligibility of audio data. Next, in terms of Spearman's rank correlation coefficient (SRCC) \cite{srcc}, our GPT-Whisper metric yields moderate correlation with human-based quality assessment (0.4360), human-based intelligibility assessment (0.5485), and Google ASR's CER (0.5415) and high correlation with Whisper's CER (0.7784). This further validates our metric's capability for zero-shot evaluation, especially given its higher correlation with CER. Next, when comparing GPT-Whisper with DNSMOS, and SpeechLMScore, GPT-Whisper excels in intelligibility metrics (0.5485 vs. 0.2643, 0.0192) while performing slightly below SpeechLMScore in quality estimation (0.5108 vs. 0.4360). Moreover, GPT-Whisper surpasses supervised non-intrusive models, including MOS-SSL and all variants of MTI-Net (wav2vec and Whisper), in predicting Whisper's CER. Experimental results show that GPT-Whisper achieves a higher SRCC (0.7784) compared to 0.7482, 0.7418, and 0.7655 for the other models.

The remainder of this paper is organized as follows. Section~\ref{sec:method} presents the proposed methodology. Section~\ref{sec:exp} describes the experimental setup and results. Finally, Section~\ref{sec:con} presents the conclusions and future work. 

\graphicspath{ {./images/} }
\begin{figure}[t]
\centering
\includegraphics[width=5cm]{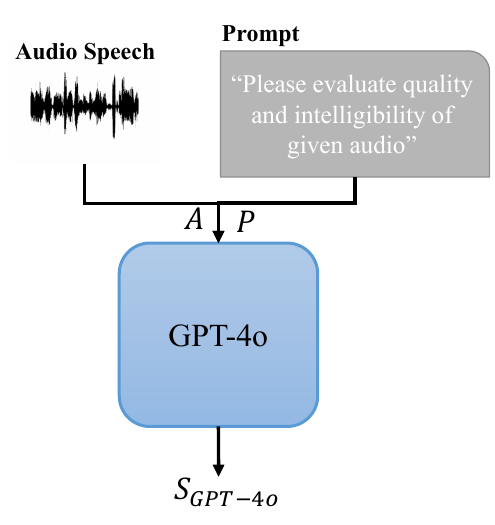} 
\caption{Zero-shot speech assessment with GPT-4o.} 
\label{fig:gpt4o}
\end{figure}

\section{Methodology}
\label{sec:method}
\subsection{GPT-4o for speech assessment}
The main intention behind using GPT-4o for speech assessment is to validate ChatGPT's reasoning capabilities in understanding speech characteristics. The overall framework for zero-shot speech assessment using GPT-4o is shown in Fig. \ref{fig:gpt4o}. Specifically, given an audio input $\textit{A}$ and a prompt $P$, ChatGPT utilizes this information to estimate the assessment metric $S_{GPT-4o}$, as follows:
\begin{equation}
\label{eq:gpt4o}
   \small
    \begin{array}{c}
    S_{GPT-4o}= ChatGPT(A,P).
    \end{array} 
\end{equation}

\graphicspath{ {./images/} }
\begin{figure}[t]
\centering
\includegraphics[width=5.3 cm]{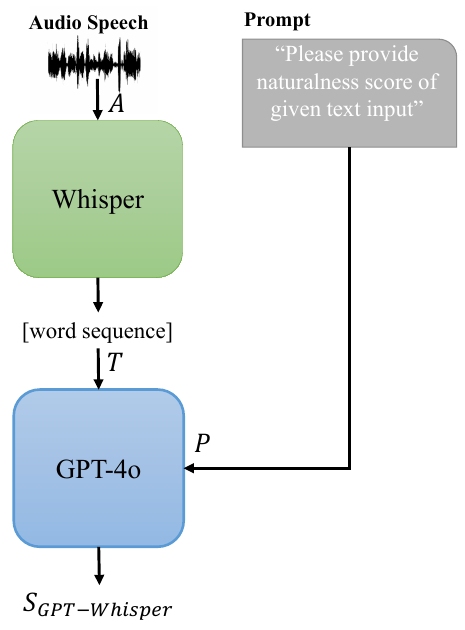} 
\caption{Zero-shot speech assessment with GPT-Whisper.} 
\label{fig:gptwhisper}
\end{figure}
\vspace{-0.5cm}
\subsection{GPT-Whisper}
The overall framework of GPT-Whisper is shown in Fig. \ref{fig:gptwhisper}. The main idea behind GPT-Whisper is to leverage the reasoning capabilities of GPT-4o to assess the predicted word sequence $\textit{T}$ from the audio input $\textit{A}$. We specifically chose Whisper \cite{Whisper} as the audio-to-text module because of its outstanding capability to convert audio into text across a variety of acoustic environments. We assume that this capability enables Whisper to capture subtle nuances of spoken language. Specifically, given an input audio $\textit{A}$, the audio-to-text conversion using Whisper ASR is defined as follows:
\begin{equation}
\label{eq:audiototext}
   \small
    \begin{array}{c}
    T= Whisper (A).
    \end{array} 
\end{equation}

Based on the predicted word sequence $\textit{T}$, we use ChatGPT to estimate the GPT-Whisper score $ S_{GPT-Whisper}$. Our strategy involves using naturalness in prompt engineering, which refers to how similar the predicted text is to human-generated text in terms of fluency, coherence, and context. This approach measures the degree to which text reflects the natural flow and nuances of human speech. The process of estimating the GPT-Whisper score is defined as follows:

\begin{equation}
\label{eq:gptwhisper}
   \small
    \begin{array}{c}
    S_{GPT-Whisper}= ChatGPT(T,P).
    \end{array} 
\end{equation}

\section{Experiments}
\label{sec:exp}
\subsection{Experimental setup}
The two proposed approaches are evaluated on the TMHINT-QI(S) dataset \cite{zezario2024studyincorporatingwhisperrobust}, which is an extension of the TMHINT-QI dataset \cite{TMINT-QI}, by including additional unseen noises, speakers, and enhancement systems. The TMHINT-QI(S) dataset is also a benchmark track in the VoiceMOS Challenge 2023 \cite{cooper2023voicemos}. The evaluation set consists of noisy, clean, and enhanced utterances, covering three seen noise conditions (babble, white, and pink) and one unseen noise condition (street). It also includes three seen speech enhancement systems (Minimum-Mean Square Error (MMSE) \cite{mmse}, Fully Convolutional Network (FCN) \cite{FCN}, and Transformer \cite{Trans}), and introduces two new, unseen speech enhancement systems: Conformer-based Metric Generative Adversarial Network (CMGAN) \cite{cao22_interspeech} and DEMUCS \cite{défossez2021music}. In total, there are 1,960 utterances with corresponding quality scores (0-5) and intelligibility scores (0-1). An additional metric, CER (0-1), is produced by two ASR models (Google ASR \cite{ASR} and Whisper \cite{Whisper}). Note that we reverse the CER score so that higher values indicate better performance, ensuring consistency with other metrics and simplifying interpretation.

Two evaluation metrics, namely linear correlation coefficient (LCC) and Spearman's rank correlation coefficient (SRCC) \cite{srcc}, are used to measure the performance of the proposed methods. Higher LCC and SRCC values indicate a stronger correlation between the predicted score and the ground-truth score, indicating better assessment prediction performance. 

\graphicspath{ {./images/} }
\begin{figure}[t]
\centering
\includegraphics[width=8.6 cm]{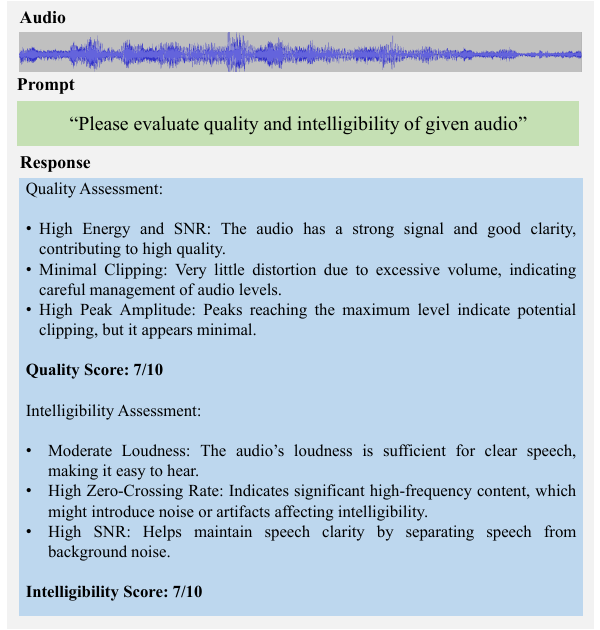} 
\caption{Example of GPT-4o used for speech assessment.} 
\label{fig:prompt}
\end{figure}

\subsection{Experimental results of GPT-4o}
In the first experiment, we intend to leverage prompt engineering on GPT-4o for speech assessment. As shown in Fig.~\ref{fig:prompt}, we specifically use keywords to evaluate the quality and intelligibility of a given audio input. The audio input in Fig.~\ref{fig:prompt} is an MMSE-enhanced speech of noisy speech containing babble noise at -2dB SNR, which is used as a representative scenario. The results show that GPT-4o determines the quality or intelligibility based only on metrics such as SNR, clipping, amplitude, or loudness. However, when we compare these results to the corresponding ground-truth information (quality score: 1.4/5, intelligibility score: 0.6/1, SNR: -2dB), GPT-4o makes inaccurate predictions. For example, as shown in Fig. \ref{fig:prompt}, GPT-4o believes that the audio input is unclipped and has a high SNR, whereas in fact the audio has a low SNR (MMSE's noise reduction effect is noticeably poor in this example) and moderate distortion. Furthermore, while the intelligibility estimate is fairly close to the true value (7/10 vs 0.6/1), the quality estimate is quite off (7/10 vs 1.4/5). Experimental results on the complete test set covering different audio characteristics show that it is difficult for the current GPT-4o to precisely predict the quality and intelligibility of the audio input. Therefore, the current GPT-4o may not serve as a reliable zero-shot non-intrusive speech assessment method.

\subsection{Experimental results of GPT-Whisper}
In this section, we aim to evaluate the capabilities of our proposed GPT-Whisper for performing zero-shot non-intrusive speech assessment.

\subsubsection{Correlation analysis}
\begin{figure}[t]
\centering
\includegraphics[width=8.6 cm]{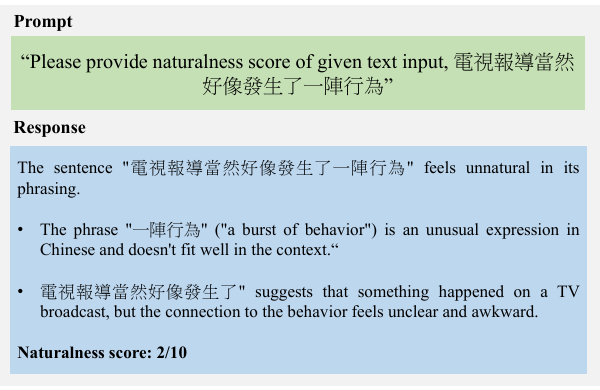} 
\caption{Example of GPT-Whisper used for speech assessment.} 
\label{fig:prompt_gpt}
\end{figure}

In initial experiments, we estimate GPT-Whisper scores using the approach outlined in Section~\ref{sec:method}. Experimental results show that GPT-4o exhibits notable reasoning capabilities in text comprehension and demonstrates solid contextual understanding, as shown in Fig. \ref{fig:prompt_gpt}. GPT-4o effectively evaluates the coherence of text inputs and determines whether connections are natural or unnatural. 

Given the above results, we aim to further investigate which assessment metrics are more correlated with the GPT-Whisper score. In our evaluation, we compare GPT-Whisper scores with human-based quality and intelligibility scores, as well as the CER of two ASR models (Whisper and Google ASR). We also deployed another variant of GPT-Whisper, namely GPT-Google, where a different ASR module is used. The results in Table \ref{tab:first} confirm that the GPT-Whisper metric achieves better overall performance than GPT-Google across nearly all metrics, emphasizing the critical role of ASR module robustness in accurate estimation. Furthermore, GPT-Whisper shows a higher correlation with intelligibility-based metrics than with quality metrics, and it has the higher correlation with the CER of Whisper ASR. Finally, when compared with other models such as DNSMOS \cite{dnsmos} and SpeechLMScore \cite{speechlm}, GPT-Whisper excels in intelligibility metrics, performs slightly below SpeechLMScore in quality estimation, and outperforms DNSMOS across all evaluation metrics.


\graphicspath{ {./images/} }

\begin{table}[t]
\caption{Comparison GPT-Whisper with other methods for estimating speech quality, speech intelligibility, Google's CER, and Whisper's CER.}
\label{tab:first}
\footnotesize
\begin{center}
\begin{tabular}{c||c|c||c|c} 
 \hline
 \hline
\multirow{2}{*}{\textbf{Model}} 
& \multicolumn{2}{c||}{\textbf{Speech Quality}} 
& \multicolumn{2}{c}{\textbf{Speech Intelligibility}} \\ 
\cline{2-5}
& \textbf{LCC} & \textbf{SRCC} & \textbf{LCC} & \textbf{SRCC} \\ 
 \hline
 \hline
GPT-Whisper         & 0.4303 & 0.4360 & \textbf{0.5226} & \textbf{0.5485} \\ 
GPT-Google (ours)   & 0.4188 & 0.4180 & 0.4641 & 0.4857 \\ 
DNSMOS \cite{dnsmos}              & 0.3088 & 0.2922 & 0.0066 & 0.0192 \\ 
SpeechLMScore \cite{speechlm}       & \textbf{0.5480} & \textbf{0.5108} & 0.2741 & 0.2643 \\ 
 \hline
 \hline
\multirow{2}{*}{} 
& \multicolumn{2}{c||}{\textbf{CER Google}} 
& \multicolumn{2}{c}{\textbf{CER Whisper}} \\ 
\cline{2-5}
GPT-Whisper         & 0.5469 & 0.5414 & \textbf{0.7541} & \textbf{0.7784} \\ 
GPT-Google (ours)   & \textbf{0.6277} & \textbf{0.6962} & 0.4635 & 0.4778 \\ 
SpeechLMScore \cite{speechlm}       & 0.2460 & 0.2070 & 0.1520 & 0.1126 \\ 
 \hline
 \hline
\end{tabular}
\end{center}
\end{table}


\subsubsection{Comparison with supervised methods}

We compare GPT-Whisper with two supervised speech assessment models, MOS-SSL \cite{ssl-mos} and MTI-Net \cite{zezario2022mti}, for predicting Whisper's CER. Specifically, Whisper's CER is used as the ground truth to evaluate the predictive performance of GPT-Whisper and the comparative assessment models. For the two comparative models, MOS-SSL incorporates wav2vec 2.0 to predict MOS scores, while MTI-Net employs cross-domain features (raw waveform, spectral features, and SSL features) to predict intelligibility metrics (intelligibility, CER, and STOI) using multi-task learning \cite{zezario2022mti}. We trained MOS-SSL and MTI-Net using the TMHINT-QI(s) training set. MOS-SSL was trained to predict Whisper's CER, while MTI-Net was trained to predict intelligibility, Google ASR's CER, Whisper's CER, and STOI. Given the significant efficacy of integrating Whisper representations into robust speech assessment models \cite{zezario2024studyincorporatingwhisperrobust, 10447907, mogridge2024nonintrusive}, we also integrated Whisper into the MTI-Net model. Therefore, we prepared two versions: MTI-Net (wav2vec) and MTI-Net (Whisper).

The results in Table~\ref{tab:1} confirm that GPT-Whisper performs comparably to supervised non-intrusive speech assessment models. Interestingly, GPT-Whisper outperforms MOS-SSL and MTI-Net (wav2vec) in LCC and SRCC. Additionally, GPT-Whisper has higher SRCC than MTI-Net (Whisper), although its LCC is slightly lower. These results demonstrate the capability of GPT-Whisper as a zero-shot speech assessment model, showing comparable or even superior performance to supervised methods.

\begin{table}[t]
\caption{Comparison of GPT-Whisper with supervised methods for estimating Whisper's CER.}
\label{tab:1}
\footnotesize
\begin{center}
 \begin{tabular}{c||c||c||c} 
 \hline
 \hline
 \textbf{System} & \textbf{Supervised} & \textbf{LCC} & \textbf{SRCC}  \\ [0.5ex] 
 \hline
\hline
GPT-Whisper&No&0.7541&\textbf{0.7784} \\  
MOS-SSL \cite{ssl-mos}&Yes&0.7328&0.7482 \\ 
MTI-Net (wav2vec) \cite{zezario2022mti}&Yes&0.7344&0.7418 \\ 
MTI-Net (Whisper) \cite{zezario2022mti}&Yes&\textbf{0.8194}&0.7655 \\ 
\hline

\end{tabular}
\end{center}
\end{table}

\graphicspath{ {./images/} }
\begin{figure}[t]
\centering
\includegraphics[width=8 cm]{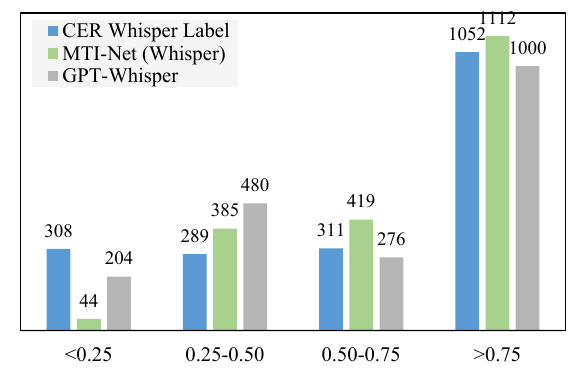} 
\caption{Distribution of predicted CER scores of GPT-Whisper and MTI-Net (Whisper) and Whisper's ground-truth CER scores.} 
\label{fig:distri}
\end{figure}

\subsubsection{Score distribution analysis}
In this section, we further analyze the distribution of scores evaluated by GPT-Whisper and MTI-Net (Whisper) and the corresponding CER labels of Whisper. This analysis aims to evaluate whether GPT-Whisper provides accurate score estimations across the entire score range. As shown in Fig.~\ref{fig:distri}, GPT-Whisper successfully predicts the entire range of scores and shows a more similar distribution to that of CER generated by the Whisper ASR, especially for CER less than 0.25. 
The closeness of the predicted CER score distribution to the true CER score distribution validates GPT-Whisper as a new assessment metric and demonstrates the potential of using a large language model for speech assessment through the GPT-Whisper method.

\section{Conclusions}
\label{sec:con}
In this paper, we have proposed two strategies for zero-shot non-intrusive speech assessment. The first strategy leverages the audio analysis capabilities of GPT-4o. The second strategy introduces a more advanced approach that uses Whisper as an audio-to-text module and evaluates the naturalness of the generated text through targeted prompt engineering, so the model is called GPT-Whisper. To the best of our knowledge, this is the first attempt to leverage ChatGPT to simultaneously estimate speech quality, intelligibility, and CER. Experimental results confirm that GPT-4o alone is insufficient for accurate speech assessment. In contrast, GPT-Whisper demonstrates potential as a zero-shot speech assessment method, with yielding moderate correlation with human-based quality assessment (SRCC of 0.4360), human-based intelligibility assessment (SRCC of 0.5485), and Google ASR's CER (SRCC of 0.5415) and notably high correlation with Whisper's CER (SRCC of 0.7784). Furthermore, compared to SpeechLMScore and DNSMOS, GPT-Whisper excels in intelligibility metrics and performs slightly below SpeechLMScore in quality estimation. Finally, in predicting Whisper’s CER, GPT-Whisper outperforms the supervised models MOS-SSL, MTI-Net (wav2vec), and MTI-Net (Whisper) in terms of SRCC (0.7784 vs. 0.7482, 0.7418, and 0.7655). This confirms GPT-Whisper's capability as a zero-shot speech assessment method and highlights the potential of large language models for advancing speech evaluation methods. In the future, we plan to explore more advanced prompt engineering to further leverage large language models to estimate other speech assessment metrics. We also intend to investigate direct integration of GPT-Whisper with speech-processing applications to enhance their performance.
 
\bibliographystyle{IEEEbib}
\bibliography{refs}

\begin{thebibliography}{10}

\bibitem{loizou2007speech}
P.~C. Loizou,
\newblock {\em Speech enhancement: theory and practice},
\newblock CRC press, 2007.

\bibitem{ref_19}
A.~Rix, J.~Beerends, M.~Hollier, and A.~Hekstra,
\newblock ``Perceptual evaluation of speech quality ({PESQ}), an objective method for end-to-end speech quality assessment of narrow-band telephone networks and speech codecs,''
\newblock in {\em ITU-T Recommendation}, 2001, p. 862.

\bibitem{katehaspi}
J.~M. Kates and K.~H. Arehart,
\newblock ``The hearing-aid speech perception index ({HASPI}) version 2,''
\newblock {\em Speech Communication}, vol. 131, pp. 35--46, 2021.

\bibitem{kates2014hearingb}
J.~M. Kates and K.~H. Arehart,
\newblock ``The hearing-aid speech quality index ({HASQI}) version 2,''
\newblock {\em Journal of the Audio Engineering Society}, vol. 62, no. 3, pp. 99--117, 2014.

\bibitem{chiang2023multiobjective}
H.-T. Chiang, S.-W. Fu, H.-M. Wang, Y.~Tsao, and J.~H.~L. Hansen,
\newblock ``Multi-objective non-intrusive hearing-aid speech assessment model,''
\newblock {\em arXiv:2311.08878}, 2023.

\bibitem{polqa_2013}
J.G Beerends, C.~Schmidmer, J.~Berger, M.~Obermann, R.~Ullmann, J.~Pomy, and M.~Keyhl,
\newblock ``Perceptual objective listening quality assessment ({POLQA}), the third generation {ITU-T} standard for end-to-end speech quality measurement part i—temporal alignment,''
\newblock {\em Journal of The Audio Engineering Society}, vol. 61, no. 6, pp. 366--384, 2013.

\bibitem{yang22o_interspeech}
Z.~Yang, W.~Zhou, C.~Chu, S.~Li, R.~Dabre, R.~Rubino, and Y.~Zhao,
\newblock ``Fusion of self-supervised learned models for {MOS} prediction,''
\newblock in {\em Proc. Interspeech}, 2022, pp. 5443--5447.

\bibitem{10447907}
S.~Cuervo and R.~Marxer,
\newblock ``Speech foundation models on intelligibility prediction for hearing-impaired listeners,''
\newblock in {\em Proc. ICASSP}, 2024, pp. 1421--1425.

\bibitem{mogridge2024nonintrusive}
R.~Mogridge, G.~Close, R.~Sutherland, T.~Hain, J.~Barker, S.~Goetze, and A.~Ragni,
\newblock ``Non-intrusive speech intelligibility prediction for hearing-impaired users using intermediate {ASR} features and human memory models,''
\newblock in {\em Proc. ICASSP}, 2024, pp. 306--310.

\bibitem{candy}
C.~O. Mawalim, B.~A. Titalim, S.~Okada, and M.~Unoki,
\newblock ``Non-intrusive speech intelligibility prediction using an auditory periphery model with hearing loss,''
\newblock {\em Applied Acoustics}, vol. 214, pp. 109663, 2023.

\bibitem{zezario2024studyincorporatingwhisperrobust}
R.~E. Zezario, Yu-Wen Chen, Szu-Wei Fu, Yu~Tsao, Hsin-Min Wang, and Chiou-Shann Fuh,
\newblock ``A study on incorporating {Whisper} for robust speech assessment,''
\newblock in {\em Proc. ICME}, 2024.

\bibitem{manocha}
P.~Manocha, D.~Williamson, and A.~Finkelstein,
\newblock ``Corn: Co-trained full- and no-reference speech quality assessment,''
\newblock in {\em Proc. ICASSP}, 2024, pp. 376--380.

\bibitem{wang2024enablingauditorylargelanguage}
S.~Wang, W.~Yu, Y.~Yang, C.~Tang, Y.~Li, J.~Zhuang, X.~Chen, X.~Tian, J.~Zhang, G.~Sun, L.~Lu, and C.~Zhang,
\newblock ``Enabling auditory large language models for automatic speech quality evaluation,''
\newblock {\em arXiv 2409.16644}, 2024.

\bibitem{cooper2024review}
E.~Cooper, W.-C. Huang, Y.~Tsao, H.-M. Wang, T.~Toda, and J.~Yamagishi,
\newblock ``A review on subjective and objective evaluation of synthetic speech,''
\newblock {\em Acoustical Science and Technology}, pp. e24--12, 2024.

\bibitem{fu2024selfsupervised}
S.-W. Fu, K.-H. Hung, Y.~Tsao, and Y.-C.~F. Wang,
\newblock ``Self-supervised speech quality estimation and enhancement using only clean speech,''
\newblock in {\em International Conference on Learning Representations}, 2024.

\bibitem{ldnet2022}
W.-C. Huang, E.~Cooper, J.~Yamagishi, and T.~Toda,
\newblock ``{LDNet}: Unified listener dependent modeling in {MOS} prediction for synthetic speech,''
\newblock in {\em Proc. ICASSP}, 2022, pp. 896--900.

\bibitem{ssl-mos}
E.~Cooper, W.-H. Huang, T.~Toda, and J.~Yamagishi,
\newblock ``Generalization ability of {MOS} prediction networks,''
\newblock in {\em Proc. ICASSP}, 2022.

\bibitem{mosa-net}
R.E Zezario, S.-W Fu, F.~Chen, C.-S Fuh, H.-M. Wang, and Y.~Tsao,
\newblock ``Deep learning-based non-intrusive multi-objective speech assessment model with cross-domain features,''
\newblock {\em IEEE/ACM Transactions on Audio, Speech, and Language Processing}, vol. 31, pp. 54--70, 2023.

\bibitem{speechlm}
S.~Maiti, Y.~Peng, T.~Saeki, and S.~Watanabe,
\newblock ``Speechlmscore: Evaluating speech generation using speech language model,''
\newblock in {\em Proc. ICASSP 2023}, 2023, pp. 1--5.

\bibitem{zezario2023multitask}
R.~E. Zezario, B.-R.~Brian Bai, C.-S. Fuh, H.-M. Wang, and Y.~Tsao,
\newblock ``Multi-task pseudo-label learning for non-intrusive speech quality assessment model,''
\newblock in {\em Proc. ICASSP}, 2024, pp. 831--835.

\bibitem{wei2024chatiezeroshotinformationextraction}
X.~Wei, X.~Cui, N.~Cheng, X.~Wang, X.~Zhang, S.~Huang, P.~Xie, J.~Xu, Y.~Chen, M.~Zhang, Y.~Jiang, and W.~Han,
\newblock ``Chatie: Zero-shot information extraction via chatting with chatgpt,''
\newblock {\em arXiv:{2302.10205}}, 2024.

\bibitem{Jia_2024_CVPR}
S.~Jia, R.~Lyu, K.~Zhao, Y.~Chen, Z.~Yan, Y.~Ju, C.~Hu, X.~Li, B.~Wu, and S.~Lyu,
\newblock ``Can chatgpt detect deepfakes? a study of using multimodal large language models for media forensics,''
\newblock in {\em Proc. of the IEEE/CVF CVPR Workshops}, 2024, pp. 4324--4333.

\bibitem{Huang_Li_Yang_Shi_Chang_Ye_Wu_Hong_Huang_Liu_Ren_Zou_Zhao_Watanabe_2024}
R.~Huang, M.~Li, D.~Yang, J.~Shi, X.~Chang, Z.~Ye, Y.~Wu, Z.~Hong, J.~Huang, J.~Liu, Y.~Ren, Y.~Zou, Z.~Zhao, and S.~Watanabe,
\newblock ``{AudioGPT}: Understanding and generating speech, music, sound, and talking head,''
\newblock {\em Proc. the AAAI Conference on Artificial Intelligence}, vol. 38, no. 21, pp. 23802--23804, 2024.

\bibitem{Whisper}
A.~Radford, J.~W. Kim, T.~Xu, G.~Brockman, C.~McLeavey, and I.~Sutskever,
\newblock ``Robust speech recognition via large-scale weak supervision,''
\newblock in {\em Proc. ICML}, 2023, pp. 28492--28518.

\bibitem{ASR}
A.~Zhang,
\newblock ``Speech recognition (version 3.6) [software], available: https://github.com/uberi/speech\_recognition\#readme,''
\newblock in {\em Proc. ICCC}, 2017.

\bibitem{dnsmos}
C.~K.~A. Reddy, V.~Gopal, and R.~Cutler,
\newblock ``{DNSMOS}: A non-intrusive perceptual objective speech quality metric to evaluate noise suppressors,''
\newblock in {\em Proc. ICASSP}, 2021, pp. 6493--6497.

\bibitem{zezario2022mti}
R.~E. Zezario, S.-W. Fu, F.~Chen, C.~S. Fuh, H.-M. Wang, and Y.~Tsao,
\newblock ``{MTI-Net}: A multi-target speech intelligibility prediction model,''
\newblock in {\em Proc. Interspeech}, 2022, pp. 5463--5467.

\bibitem{srcc}
C.~Spearman,
\newblock ``The proof and measurement of association between two things,''
\newblock {\em The American Journal of Psychology}, vol. 15, no. 1, pp. 72--101, 1904.

\bibitem{TMINT-QI}
Y.-W. Chen and Y.~Tsao,
\newblock ``{InQSS}: a speech intelligibility assessment model using a multi-task learning network,''
\newblock in {\em Proc. Interspeech}, 2022, pp. 3088--3092.

\bibitem{cooper2023voicemos}
E.~Cooper, W.-C. Huang, Y.~Tsao, H.-M. Wang, T.~Toda, and J.~Yamagishi,
\newblock ``{The VoiceMOS Challenge 2023}: Zero-shot subjective speech quality prediction for multiple domains,''
\newblock in {\em Proc. ASRU}, 2023, pp. 1--7.

\bibitem{mmse}
Y.~Ephraim and D.~Malah,
\newblock ``Speech enhancement using a minimum mean-square error log-spectral amplitude estimator,''
\newblock {\em IEEE Transactions on Acoustics, Speech, and Signal Processing}, vol. 33, no. 2, pp. 443--445, 1985.

\bibitem{FCN}
S.-W. Fu, Y.~Tsao, X.~Lu, and H.~Kawai,
\newblock ``Raw waveform-based speech enhancement by fully convolutional networks,''
\newblock in {\em Proc. APSIPA ASC}, 2017.

\bibitem{Trans}
J.~Kim, M.~El-Khamy, and J.~Lee,
\newblock ``{T-GSA}: Transformer with {Gaussian}-weighted self-attention for speech enhancement,''
\newblock in {\em Proc. ICASSP}, 2020, pp. 6649--6653.

\bibitem{cao22_interspeech}
R.~Cao, S.~Abdulatif, and B.~Yang,
\newblock ``{CMGAN}: Conformer-based {Metric GAN} for speech enhancement,''
\newblock in {\em Proc. Interspeech 2022}, 2022, pp. 936--940.

\bibitem{défossez2021music}
A.~Défossez, N.~Usunier, L.~Bottou, and F.~Bach,
\newblock ``Music source separation in the waveform domain,''
\newblock {\em arXiv 1911.13254}, 2021.

\end{thebibliography}
\end{document}